\documentclass[conference,a4paper]{APSIPA2021}

\usepackage{amsmath}
\usepackage{graphicx}
\usepackage{multirow}
\usepackage{threeparttable}
\usepackage{url} 
\usepackage{amsfonts}

\usepackage{geometry}
\usepackage{fancyhdr}

\usepackage{booktabs}
\usepackage{multirow}

\usepackage{CJK}

\title{Can We Estimate Purchase Intention Based on\\Zero-shot Speech Emotion Recognition?}

\author{
\authorblockN{
Ryotaro Nagase\authorrefmark{1}\authorrefmark{2}, 
Takashi Sumiyoshi\authorrefmark{1},
Natsuo Yamashita\authorrefmark{1},
Kota Dohi\authorrefmark{1}, and
Yohei Kawaguchi\authorrefmark{1}
}
\authorblockA{
\authorrefmark{1}
Hitachi, Ltd., Japan \\
E-mail: yohei.kawaguchi.xk@hitachi.com}
\authorblockA{
\authorrefmark{2}
Ritsumeikan University, Japan}
}

\begin{document}
\maketitle

\begin{abstract}
This paper proposes a zero-shot speech emotion recognition (SER) method that estimates emotions not previously defined in the SER model training. Conventional methods are limited to recognizing emotions defined by a single word. Moreover, we have the motivation to recognize unknown bipolar emotions such as ``I want to buy - I do not want to buy.'' In order to allow the model to define classes using sentences freely and to estimate unknown bipolar emotions, our proposed method expands upon the contrastive language-audio pre-training (CLAP) framework by introducing multi-class and multi-task settings. We also focus on purchase intention as a bipolar emotion and investigate the model's performance to zero-shot estimate it. This study is the first attempt to estimate purchase intention from speech directly. Experiments confirm that the results of zero-shot estimation by the proposed method are at the same level as those of the model trained by supervised learning.
\end{abstract}


\section{Introduction}
Speech emotion recognition (SER) is a technique to estimate emotions conveyed by speech.
This technique can be applied to the speech analysis of call centers \cite{app_call_center}, spoken dialogue agent \cite{app_alexa}, e-learning system \cite{app_e_learning}, and mental health analysis \cite{app_mental_health}, etc.

Many researchers have proposed various methods to improve the performance of SER.
Especially, in recent years, there are many methods using deep learning such as SER with self-supervised learning model \cite{SER_SSL1, SER_SSL2}.
In these methods, the model of SER is often trained by supervised-learning.
Therefore, the trained model can recognize predefined seen emotions, while it is difficult to recognize unseen emotions.
To alleviate this limitation, previous studies have proposed methods of zero-shot SER that can recognize undefined emotions.
For example, X. Xu, et al. proposed the method using auditory affective descriptors (AAD) to transfer knowledge from the seen domain to the unseen domain \cite{xu19_interspeech}, and the method using reconstructed semantic prototypes and data augmentation for the unseen domain \cite{Xu23_icassp}.

The challenge of zero-shot SER is how to define the classes of emotions.
Conventional methods do not assume the estimation of emotion classes that cannot be defined by a single word.
Therefore, it is difficult to recognize the emotions expressed by text, such as ``the feeling of willing to buy (purchase intention).''
In order to realize SER capable of zero-shot recognizing various emotions, it is necessary to have a framework that can freely define classes during the training and estimation phase.
We also have the motivation to realize the zero-shot estimation of bipolar emotions, such as ``I want to buy - I do not want to buy.''

To meet these requirements, in this study, we propose a new contrastive language-audio pre-training (CLAP) method that can perform zero-shot estimation of bipolar emotions. 
CLAP can decide the emotion freely because it can represent the classification class by the sentence.
In our method, we define the multi-class emotions with a bipolar sub-class and train the model with CLAP for each emotion.
We use six basic emotions with a bipolar sub-class as the multi-class emotions.
We expect this model trained by the proposed method to correctly classify the emotions using the knowledge of six basic emotions, even if the estimated class is an unknown bipolar emotion. 
We also focus on purchase intention as a bipolar emotion and experiment on whether the model trained by the proposed method can zero-shot estimate purchase intention.
The contributions of this study are as follows.
\begin{itemize}
    \item This is the first method for zero-shot SER capable of estimating bipolar emotions, such as purchase intention.
    \item This study is the first time to directly estimate purchase intention from speech.
\end{itemize}
This paper proceeds as follows.
In Section2, we describe the method of the basic CLAP, proposed method with CLAP, and the data augmentation. In Section3, we explain the setup for the experiment and show the results. Finally, in Section4, we present our conclusions and future work.

\section{Methodology}
\subsection{CLAP}
CLAP is the training scheme that calculates the similarity or dissimilarity between acoustic and linguistic embeddings \cite{Elizalde22_CLAP}.
This method makes it possible for us to define the categories during the estimation phase and classify speech into unseen categories.
Moreover, the model can estimate more various categories because CLAP can define them with the sentence. 

\begin{figure}[t]
    \centering
    \includegraphics[scale=0.5]{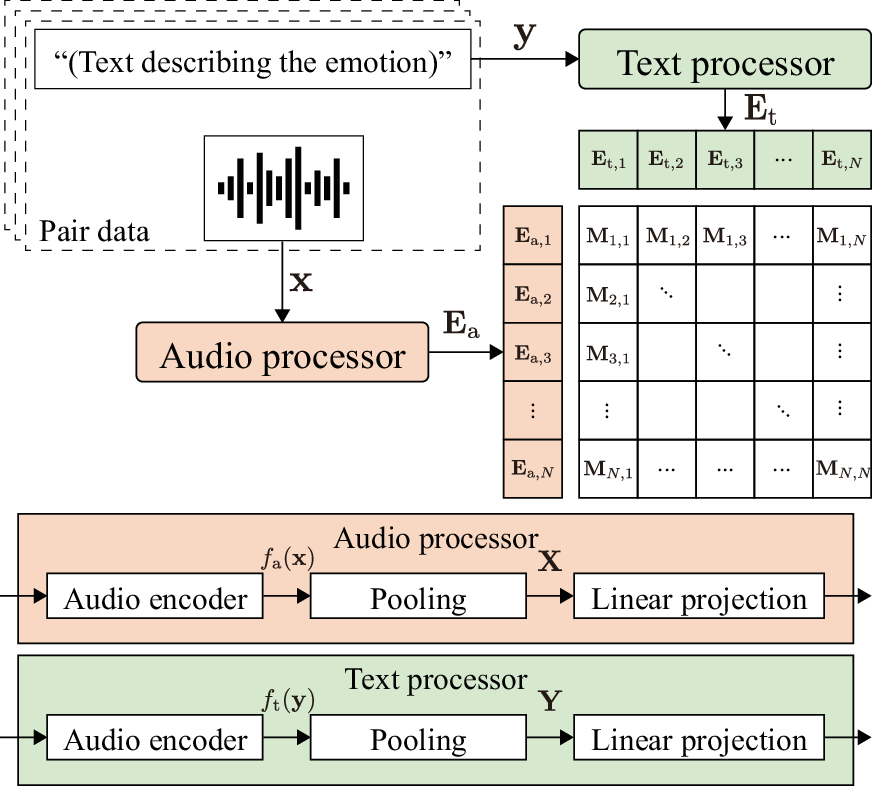}
    \caption{Overview of clap during the training phase}
    \label{fig:overview_clap_training}
\end{figure}

\begin{figure}[t]
    \centering
    \includegraphics[scale=0.5]{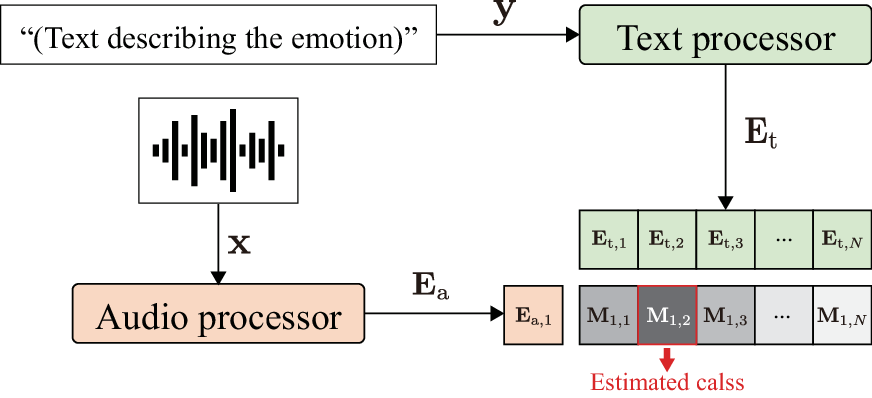}
    \caption{Overview of clap during the estimation phase}
    \label{fig:overview_clap_estimating}
    \vspace{-1.0em}
\end{figure}

The overviews of CLAP during the training or estimation phase are shown in Figures~\ref{fig:overview_clap_training} and~\ref{fig:overview_clap_estimating}, respectively.
The inputs are audio and text of categories. 
Let $\mathbf{x} \in \mathbb{R}^{N \times T}$, $\mathbf{y} \in \mathbb{R}^{N \times L}$ be the audio and the text, respectively, where $T$ and $L$ are the length of them, and $N$ is the size of mini-batch.
Note that $\mathbf{x}$, $\mathbf{y}$ have correspondence in mini-batch.

The inputs are converted to the embeddings by the audio and text encoders.
The audio encoder $f_{\rm a}(\cdot)$ and text encoder $f_{\rm t}(\cdot)$ are given in Equation \ref{eq:encoder_audio} and \ref{eq:encoder_text}, respectively.
Let $\mathbf{X} \in \mathbb{R}^{N \times D_{\rm a}}$, $\mathbf{Y} \in \mathbb{R}^{N \times D_{\rm t}}$ be the audio embeddings and text embeddings, where $D_{a}$, $D_{t}$ are the number of dimensions of each encoder's output.
Moreover, $\rm Pooling$ is applied to average or select the embeddings from each encoder.
\begin{eqnarray}
    \mathbf{X} &=& {\rm Pooling}(f_{\rm a}(\mathbf{x})) \label{eq:encoder_audio}\\
    \mathbf{Y} &=& {\rm Pooling}(f_{\rm t}(\mathbf{y})) \label{eq:encoder_text}
\end{eqnarray}
Those embeddings are input into the linear projection, which are given in Equation \ref{eq:fixed_audio} and \ref{eq:fixed_text}.
Let $\mathbf{E}_{\rm a} \in \mathbb{R}^{N \times D}$, $\mathbf{E}_{\rm t} \in \mathbb{R}^{N \times D}$ be the fixed embeddings of $D$ dimensions from each linear projection.
Also, $\rm Linear_{a} (\cdot)$, $\rm Linear_{t} (\cdot)$ are the linear projection for audio and text, respectively.
\begin{eqnarray}
    \mathbf{E}_{\rm a} &=& {\rm Linear_{a}}(\mathbf{X}) \label{eq:fixed_audio}\\
    \mathbf{E}_{\rm t} &=& {\rm Linear_{t}}(\mathbf{Y}) \label{eq:fixed_text}
\end{eqnarray}

During the training phase, we calculate the similarity matrix $\mathbf{M} \in \mathbb{R}^{N \times N}$ given in Equation \ref{eq:similarity}.
Let $\tau$ be the temperature parameter. 
\begin{eqnarray}
    \mathbf{M} &=& \tau(\mathbf{E}_{\rm t} \cdot \mathbf{E}_{\rm a}^{\top}) \label{eq:similarity}
\end{eqnarray}
After that, we calculate and optimize the symmetric loss function \cite{Radford21_CLIP} $\mathcal{L}(\cdot)$ given in Equation \ref{eq:loss_all}, where ${\rm CE(\cdot)}$ is the function of cross entropy, and $\hat{\mathbf{M}} \in \mathbb{R}^{N \times N}$ is the ground truth matrix. 
\begin{eqnarray}
    \mathcal{L}(\mathbf{M}, \hat{\mathbf{M}}) &=& \frac{1}{2}({\rm CE}(\mathbf{M}, \hat{\mathbf{M}})+{\rm CE}(\mathbf{M}^{\top}, \hat{\mathbf{M}})) \label{eq:loss_all}
\end{eqnarray}
Optimizing this loss function, we can train the model to minimize the loss between the ground truth and the estimated values and maximize the loss between the incorrect answer and the estimated values.

% In estimating phase, we define the categories of the classification such as words or sentences.
During the estimation phase, we defined the categorical classes by words or sentences.
Then, we input the speech and texts represented the categories into the encoder and linear projection corresponding to each modalities, and obtain fixed-length vectors.
Finally, the similarity between speech and text is calculated using the fixed-length vectors, and the class indicated by the text of the highest similarity is the estimation result.

This method has been applied to SER.
For example, in the original paper on CLAP, zero-shot SER was performed using the CLAP model trained with environmental sound datasets \cite{Elizalde22_CLAP}.
Y. Pan et al. also proposed GEmo-CLAP for gender and emotion classification and improved the accuracy rate of four emotion classifications that were not zero-shot \cite{pan23_gemoclap}.
However, there is still no research on recognizing emotion-related classes such as purchase intention using the model of zero-shot SER trained with emotional speech datasets.

\subsection{Zero-shot SER with multi-class multi-task CLAP}
\begin{figure}
    \centering
    \includegraphics[scale=0.5]{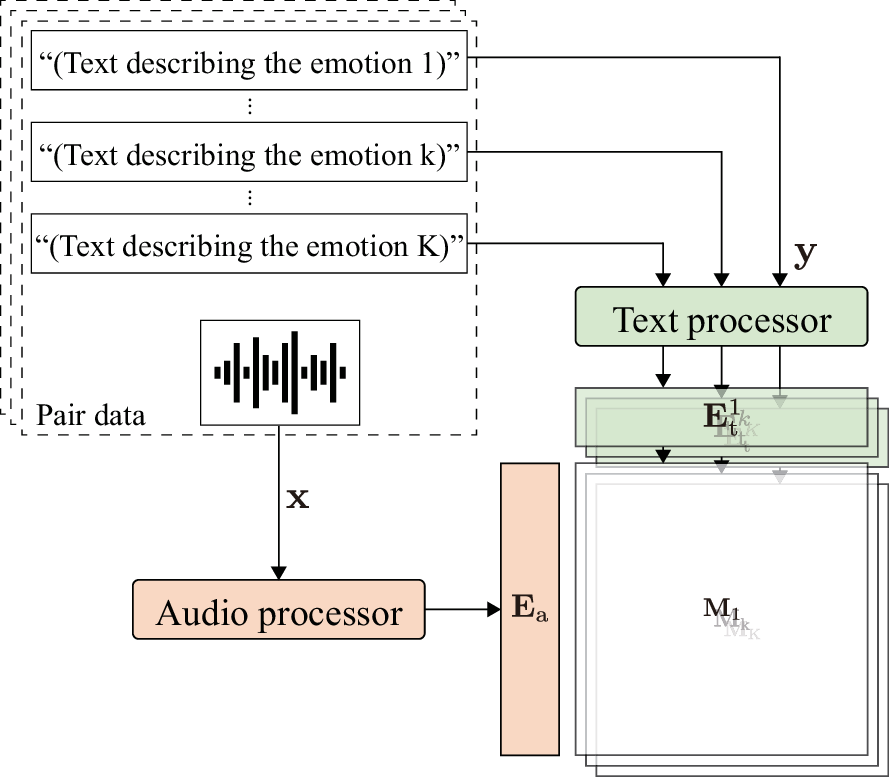}
    \caption{Overview of the proposed method}
    \label{fig:overview_mmclap}
    \vspace{-1.0em}
\end{figure}

We propose the training method of the zero-shot SER for multi-class emotional speeches, which is called multi-class multi-task CLAP.
The overview of the proposed method is shown in Figure~\ref{fig:overview_mmclap}.

During the training phase, the model outputs the similarity matrix of the multi-class prepared for each emotion. 
After calculating the contrastive loss with the ground truth matrix in each emotion, we summarize and optimize them.
Equation \ref{eq:mmclap} formulates the calculation of the loss $\mathcal{L}_{\rm all}$.
Let $K$ be the number of emotion classes, $k$ be the index of an emotion class, and $\mathbf{M}_{k} \in \mathbb{R}^{N \times N}$, $\hat{\mathbf{M}}_{k} \in \mathbb{R}^{N \times N}$ be the similarity matrix and ground truth matrics in each emotion, respectively.
\begin{eqnarray}
    \mathcal{L}_{\rm all} &=& \sum_{k=1}^{K} \mathcal{L}(\mathbf{M}_{k}, \hat{\mathbf{M}}_{k}) \label{eq:mmclap}
\end{eqnarray}

During the estimation phase, we also define the categories and classify speeches as before.
The model trained by the proposed method can recognize multi-class emotions with the sub-class.
Therefore, we expect the proposed method will enable the zero-shot SER to estimate emotions such as unknown bipolar emotions.

\subsection{The data augmentation of multi-class by paraphrasing}
We extend the dataset by paraphrasing the text indicating the multi-class to increase the vocabulary and grammar of text associated with speech.
Paraphrasing was performed using GPT4, a large-scale language model provided by OpenAI\footnote{\url{https://openai.com/gpt-4}}.
The template of prompts given to GPT4 is shown in Table~\ref{tab:template_prompt}.
\begin{table}[t]
    \begin{center}
        \caption{Template of the prompt for paraphrasing}
        \label{tab:template_prompt}
        \begin{tabular}{@{}p{25em}@{}}
        \toprule
        \textbf{\# Prompt for paraphrasing}
        \\ \midrule
        You are an imaginative assistant.\\
        Given the emotion description, please generate ten paraphrase sentences in Japanese.\\
        Note that you cannot add the prohibited word.\\
        \\
        \#\#\# Input and output\\
        Emotion description: \lbrack e.g. \begin{CJK}{UTF8}{min}私の気持ちは覚醒しています\end{CJK} \\ (My emotion is aroused.) \rbrack\\
        Prohibited word: \lbrack e.g. \begin{CJK}{UTF8}{min}覚醒\end{CJK} (aroused) \rbrack\\
        Paraphrase sentences:
        \\ \bottomrule
        \end{tabular}
    \end{center}
    \vspace{-1.0em}
\end{table}
GPT4 performs better in English than Japanese \cite{OpenAI23_GPT4}, so the prompt is written in English.
The prompt includes the text of multi-class and the forbidden word.
If the forbidden word is included in the output, we modify it by hand.

\section{Experiments}
We experimented with the binary classification task of whether purchase intention from speech.
In the following, we explain the experimental setup and results.

\subsection{Dataset}
Before the experiment, we created the internal dataset which has annotated emotional utterances in Japanese.
First, we collected emotional utterances by role-playing conversations.
We prepared six scenarios related to conversations between salespeople and customers.
Five Japanese speakers talked freely for each scenario.
Note that we did not prepare sentences for reading.
The sampling rate of the recorded utterance was 44,100 Hz.
The recorded data were separated into sessions for each scenario, and we obtained the dataset consisting of six sessions.
Afterward, three annotators listened to collected utterances and scored the intensity values of the six emotions (pleasant-unpleasant, aroused-sleepy, dominant-submissive, credible-doubtful, interested-indifferent, positive-negative) on a seven-point scale (1.very low--7.very high).
These emotions were assigned according to the related work \cite{MORI11_SC}.
In addition, these annotators scored the intensity values indicating the purchase intention on a seven-point scale for only customer's data at corrected utterances.

Finally, we arranged the labels for the classification task.
The distribution of scores of each annotator was normalized by the mean and variance of the distribution of scores given by all annotators.
The normalized labels exclude the most frequent value and sets it as a threshold.
The score below the threshold are relabeled as negative labels (0) and the score above the threshold are relabeled as positive labels (1).
The number of utterances in two classes of six emotions was 2,598, and the number of utterances with the two classes of the purchase intention was 940.
We converted the sampling rate of each utterance to 16,000Hz.

During the training and estimation phase using the proposed method, we used descriptions that explain the emotions corresponding to the scores aggregated above.
Table~\ref{tab:descriptions} shows the correspondence between the class scores of each emotion and the descriptions of each emotion.
\begin{table*}[t]
    \begin{center}
        \caption{Correspondence between two classes of each emotion and the descriptions of each emotion}
        \label{tab:descriptions}
        {\renewcommand\arraystretch{1.3}
        \begin{tabular}{lll}
        \toprule
        & Negative label (0)  & Positive label (1) \\ \midrule
        Unpleasant-pleasant &
          \begin{tabular}[c]{@{}l@{}}\begin{CJK}{UTF8}{min}私は不快な気持ちです\end{CJK}\\(My emotions are uncomfortable.)\end{tabular}&
          \begin{tabular}[c]{@{}l@{}}\begin{CJK}{UTF8}{min}私は快な気持ちです\end{CJK}\\(My emotions are pleasant.)\end{tabular} \\ \hline
        Sleepy-aroused &
          \begin{tabular}[c]{@{}l@{}}\begin{CJK}{UTF8}{min}私の気持ちは睡眠しています\end{CJK}\\(My emotions are asleep.)\end{tabular} &
          \begin{tabular}[c]{@{}l@{}}\begin{CJK}{UTF8}{min}私の気持ちは覚醒しています\end{CJK}\\(My emotions are awake.)\end{tabular} \\ \hline
        Submissive-dominant &
          \begin{tabular}[c]{@{}l@{}}\begin{CJK}{UTF8}{min}私の気持ちは従属的です\end{CJK}\\(My emotions are subservient.)\end{tabular} &
          \begin{tabular}[c]{@{}l@{}}\begin{CJK}{UTF8}{min}私の気持ちは支配的です\end{CJK}\\(My emotions are dominant.)\end{tabular} \\ \hline
        Doubtful-credible & \begin{tabular}[c]{@{}l@{}}\begin{CJK}{UTF8}{min}私は不信に思います\end{CJK}\\(I distrust.)\end{tabular}&\begin{tabular}[c]{@{}l@{}}\begin{CJK}{UTF8}{min}私は信頼しています\end{CJK}\\(I trust.)\end{tabular}\\ \hline
        Indifferent-interested &
          \begin{tabular}[c]{@{}l@{}}\begin{CJK}{UTF8}{min}私は無関心です\end{CJK}\\(I am indifferent.)\end{tabular} &
          \begin{tabular}[c]{@{}l@{}}\begin{CJK}{UTF8}{min}私は関心があります\end{CJK}\\(I am interested.)\end{tabular} \\ \hline
        Negative-positive & \begin{tabular}[c]{@{}l@{}}\begin{CJK}{UTF8}{min}私は否定的です\end{CJK}\\(I am negative.)\end{tabular} & \begin{tabular}[c]{@{}l@{}}\begin{CJK}{UTF8}{min}私は肯定的です\end{CJK}\\(I am positive.)\end{tabular} \\ \bottomrule
        \end{tabular}}
    \end{center}
    \vspace{-1.0em}
\end{table*}
In addition, we experimented with the augmented data by paraphrasing, which was 11 times the number of the original dataset.
In the evaluation, we used one of the six sessions as the testing data and others as the training data.

\subsection{Models and metrics}
We explain the model structure of the proposed method.
The audio encoder consisted of Japanese HuBERT \cite{Hsu21_TASLP} and linear projection.
The output dimensions of them were 768 and 512, respectively.
Before passing embeddings into linear projection, we averaged the audio embeddings obtained from HuBERT in the time direction.
The text encoder consisted of Japanese DistilBERT \cite{Sanh19_arXiv} and linear projection.
The output dimensions of them were also 768 and 512, respectively.
Before passing embeddings into linear projection, we only used class token from the embeddings obtained from DistilBERT.
We used the pre-trained parameters of HuBERT\footnote{\url{rinna/japanese-hubert-base}} and DistilBERT\footnote{\url{laboro-ai/distilbert-base-japanese}} provided by Hugging Face \cite{wolf-etal-2020-transformers}.

The baseline in this paper was the model to estimate purchase intention by supervised learning instead of CLAP.
This model structure was a combination of HuBERT, linear projection, and one fully connected layer, and we used the pre-trained HuBERT in the same way as the proposed method.
The threshold for binary classification on the baseline is defined by the Youden index of the receiver operatorating characteristic (ROC) curve.
The number of epochs was 300, the batch size was 64, the learning rate was 0.000001, and the optimization method was Adam \cite{kingma14_adam}.
The loss functions of the baseline and the proposed method were binary cross entropy loss and symmetric cross entropy loss, respectively.
\begin{table*}[t]
    \begin{center}
        \caption{WA, UA, and the recall of purchase intention}
        \label{tab:pie}
        \scalebox{0.95}{
        \begin{tabular}{@{}clllllllll@{}}
        \toprule
        \multicolumn{1}{l}{\multirow{3}{*}{}} &
          \multirow{3}{*}{Text indicating classes (ja)} &
          \multicolumn{4}{l}{No augmentation} &
          \multicolumn{4}{l}{Augmentation by paraphrasing} \\ \cmidrule(l){3-10} 
        \multicolumn{1}{l}{} &
           &
          \multirow{2}{*}{WA} &
          \multirow{2}{*}{UA} &
          \multicolumn{2}{l}{Purchase intention} &
          \multirow{2}{*}{WA} &
          \multirow{2}{*}{UA} &
          \multicolumn{2}{l}{Purchase intention} \\ \cmidrule(lr){5-6} \cmidrule(l){9-10} 
        \multicolumn{1}{l}{} &
           &
           &
           &
          No &
          Yes &
           &
           &
          No &
          Yes \\ \midrule
        Random &
          - &
          49.5 &
          49.4 &
          48.8 &
          50.0 &
          - &
          - &
          - &
          - \\ \midrule
        \begin{tabular}[c]{@{}c@{}}Supervised\\ {[}Baseline{]}\end{tabular} &
          - &
          \textbf{74.0} &
          \textbf{69.2} &
          \textbf{78.5} &
          60.0 &
          - &
          - &
          - &
          - \\ \midrule
        \multirow{3}{*}{\begin{tabular}[c]{@{}c@{}}Zero-shot\\ {[}Ours{]}\end{tabular}} &
          \begin{tabular}[c]{@{}l@{}}(1) \begin{CJK}{UTF8}{min}私は買う気\{なし, あり\}です\end{CJK}\\ (I am \{unwilling / willing\} to buy.)\end{tabular} &
          60.2 &
          69.1 &
          51.6 &
          \textbf{86.7} &
          61.8 &
          62.3 &
          61.3 &
          \textbf{63.3} \\ \cmidrule(l){2-10} 
         &
          \begin{tabular}[c]{@{}l@{}}(2) \begin{CJK}{UTF8}{min}私は購買意欲\{なし, あり\}\end{CJK}\\ (I \{do not have / have\} the purchase intention.)\end{tabular} &
          64.2 &
          60.5 &
          67.7 &
          53.3 &
          65.0 &
          63.3 &
          66.7 &
          60.0 \\ \cmidrule(l){2-10}
         &
          \begin{tabular}[c]{@{}l@{}}(3) \begin{CJK}{UTF8}{min}私は\{欲しくない,欲しい\}です\end{CJK}\\ (I \{do not want / want\} to buy it.)\end{tabular} &
          61.0 &
          61.8 &
          60.2 &
          63.3 &
          \textbf{73.2} &
          \textbf{69.8} &
          \textbf{76.3} &
          \textbf{63.3} \\ \bottomrule
        \end{tabular}
        }
    \end{center}
    \vspace{-1.0em}
\end{table*}

We used weighted accuracy (WA), unweighted accuracy (UA), and recall of each category as the evaluation metrics.
Note that WA is the total number of correct answers to the total data, and UA is the average recall of each category.
The higher these metrics are, the more correctly the performance can estimate purchase intention. 
Moreover, we calculated the frequency rate of matching the random value and the correct answer as a chance rate.
In the evaluation of the proposed method, we used the three texts.
They all indicated the purchase intention, while the words used in the sentence differed.
We used these texts to compare the proposed method with the chance rate and baseline.

\subsection{Results}
Table~\ref{tab:pie} shows WA, UA, and recall of each category.
The result of the random was about 50\% for both WA and UA.
Comparing the result of the random with the baseline, WA was 24.5\%, and UA was 19.8\% higher in the baseline result than in the random.
These results indicate that the baseline model does not make a random estimation.
It also shows that it is possible to estimate purchase intention from speech using deep learning.
Comparing the result of the random with the proposed methods, WA and UA were higher in all proposed methods than in the random.
These results show that the zero-shot estimation by the proposed method is not random, and these methods can train the model to estimate unknown bipolar emotions.

Without data augmentation, the results show that the recall with purchase intention ``Yes'' using text (1) is higher in the proposed method than in the baseline.
The UA using text (1) and the recall with purchase intention ``Yes'' using text (3) were similar to the baseline result.
On the other hand, the recalls with purchase intention ``No'' using text (1) and ``Yes'' using text (2) were lower than 60\%.
It indicates that the model trained by the proposed method cannot represent the correspondence well between the semantic information of text and speech under some conditions.

All evaluation scores with data augmentation are improved overall than those without data augmentation.
Those scores equal the baseline, especially when using text (3).
This indicates that data augmentation make it possible for the model to train the correspondence between text and speech semantic information relatively more easy.
These results indicate that the zero-shot SER trained by the proposed method can estimate purchase intention that is unseen class by data augmentation of paraphrasing and appropriate creation of text indicating classes.

We also investigated the significant difference between the evaluation scores of the baseline and the proposed method using text (3) through a sign test.
If there was no significant difference between them, we could assume that the number of samples that were correct by the baseline and incorrect by the proposed method was the same as those that were correct by the proposed method and incorrect by the baseline.
We performed a two-tailed sign test based on the above hypothesis.
The p-value $p$ about the baseline and the proposed method using text (3) without data augmentation was less than 0.05 ($p<0.05$).
On the other hand, the p-value $p$ about the baseline and the proposed method using text (3) with data augmentation was more than 0.05 ($p>0.05$).
These results indicated that the significant difference between the baseline and the proposed method decreased with data augmentation. 
In other words, when using text (3) along with data augmentation, the proposed method can provide zero-shot estimates comparable to the supervised method.

\section{Conclusion}
This study proposed a new method for zero-shot speech emotion recognition.
The proposed method is a multi-class multi-task CLAP that can recognize unknown bipolar emotions. 
We found that zero-shot SER trained by the proposed method with paraphrasing data augmentation can be used to estimate purchase intention with the same accuracy as a supervised learning model.
In the future, we would like to expand speech data with multi-class labels and perform zero-shot inference for generously related classes other than purchase intention.

%\printbibliography

\bibliographystyle{IEEEtran}
\bibliography{mybib.bib}

\end{document}